\documentclass[aps,pra,superscriptaddress,twocolumn,showpacs,amssymb,amsfonts,
amsmath]{revtex4-1}
\usepackage{mathrsfs,bbm}
\usepackage{graphicx}
\usepackage{amsthm}
\usepackage{times}
\usepackage{color}
\usepackage{enumerate}

\definecolor{myurlcolor}{rgb}{0,0,0.7}
\definecolor{myrefcolor}{rgb}{0.8,0,0}
\usepackage{hyperref}
\hypersetup{colorlinks, linkcolor=myrefcolor,
citecolor=myurlcolor, urlcolor=myurlcolor}

\newtheorem{thm}{Theorem}
\newtheorem{lem}[thm]{Lemma}
\newtheorem{cor}[thm]{Corollary}

\usepackage[draft]{fixme}
\usepackage{changes}
\usepackage[T1]{fontenc}

\def\textbf#1{{\bf #1}}
\def\be{\begin{equation}}
\def\ee{\end{equation}}
\def\ben{\begin{eqnarray}}
\def\een{\end{eqnarray}}
\def\eea{\end{array}}
\def\bea{\begin{array}}
\newcommand{\bei}{\begin{itemize}}
\newcommand{\eei}{\end{itemize}}

\newcommand{\Tr}[0]{\mathrm{\mbox{Tr}}}
\newcommand{\ot}{\otimes}
\newcommand{\ket}[1]{|#1\rangle}
\newcommand{\bra}[1]{\langle#1|}
\newcommand{\proj}[1]{\ket{#1}\bra{#1}}

\begin{document}

\title{Separability in terms of a single entanglement witness}
\author{Piotr Badzi\k{a}g}
\affiliation{Alba Nova Fysikum, University of Stockholm,
S--106~91, Sweden}

\author{Pawe\l{} Horodecki}
\affiliation{Faculty of Applied Physics and Mathematics, Gda\'nsk
University of Technology, 80--952 Gda\'nsk, Poland}

\author{Ryszard Horodecki}
\affiliation{Institute of Theoretical Physics and Astrophysics,
University of Gda\'nsk, 80-952 Gda\'nsk, Poland}

\author{Remigiusz Augusiak}
\affiliation{ICFO--Institut de Ci\`{e}ncies Fot\`{o}niques, 08860
Castelldefels (Barcelona), Spain} \affiliation{Faculty of Applied
Physics and Mathematics, Gda\'nsk University of Technology,
80--952 Gda\'nsk, Poland}

\begin{abstract}
The separability problem is formulated in terms of a
characterization of a {\it single} entanglement witness. More
specifically, we show that any (in general multipartite) state
$\varrho$ is separable if and only if a specially constructed
entanglement witness $W_{\varrho}$ is weakly optimal, i.e., its
expectation value vanishes on at least one product vector.
Interestingly, the witness can always be chosen to be
decomposable. Our result changes the conceptual aspect of the
separability problem and rises some questions about properties
of positive maps.
\end{abstract}

\maketitle

\textit{Introduction.} One of the fundamental problems in quantum
information theory concerns detection and characterization of
entanglement. In many instances, questions concerning detection
can be successfully addressed via the theory of positive maps.
There, separability (i.e., the absence of entanglement) of
$\varrho \in B(\mathcal{H}_A \otimes \mathcal{H}_B)$ is equivalent
to the statement that for all positive maps $\Lambda$ acting on
$B(\mathcal{H}_A)$, the operator $\sigma=[\Lambda \otimes
\mathrm{id}](\varrho)$, with $\mathrm{id}$ denoting the identity
map acting on $B(\mathcal{H}_B)$, is positive \cite{Peres,sep}.
Via Jamio\l{}kowski's isomorphism \cite{JamiolkowskiIsomorphism},
the latter can be reformulated in terms of physical (Hermitian)
operators instead of positive maps \cite{sep}. Precisely, the
state $\varrho$ is separable if and only if the following non-negativity
condition
\begin{equation}\label{condition}
\langle W\rangle_{\varrho}:=\Tr(W \varrho)\geq 0
\end{equation}
is satisfied for every Hermitian operator $W \in B(\mathcal{H}_A
\otimes \mathcal{H}_B)$ such that (a) $\langle \alpha,\beta
|W|\alpha,\beta \rangle \geq 0$ for all product vectors
$|\alpha,\beta\rangle\equiv \ket{\alpha}\otimes\ket{\beta} \in
\mathcal{H}_A \otimes \mathcal{H}_B$, and (b) there is an entangled
state $\sigma\in B(\mathcal{H}_A \otimes \mathcal{H}_B)$ for which
$\langle W\rangle_{\sigma}<0$.

The importance of this formulation was first recognized by Terhal
\cite{Terhal} (see also Ref. \cite{Exp}), who coined the term
\textit{entanglement witness} for these operators. Also, Terhal
pointed out the possibility of experimental entanglement tests via
verification of the condition (\ref{condition}) in a laboratory.
Since then entanglement witnesses have become one of the most
popular tools for entanglement detection, as they allow to
identify entanglement without otherwise difficult to avoid
complete state tomography \cite{Witn1} (for nonlinear and other
methods of entanglement detection see, e.g., Ref. \cite{CovarianceSep}
and also the recent reviews \cite{przegladowa,przeglDet}). Owing
to this, entanglement witnesses have been a subject of rigorous
studies leading to a better understanding of their properties and
numerous methods of construction (see, e.g., Refs.
\cite{Optimisation,swiadki:wlasn,przegladowa,przeglDet}). More
importantly, their impressive experimental implementations have
been performed \cite{Experiments}.

Despite all the progress, practical characterization of the set of
entanglement witnesses, which would provide precise optimization
parameters is still eluding the researches. Usually, the
parameters can only be estimated with limited accuracy
\cite{Exp,Optimisation} and the entanglement witnesses have a
structure, which is not easy to handle.

As part of the effort to improve on this unsatisfactory situation,
in this paper we simplify the conceptual aspect of the
separability problem at a cost of the size of the underlying
Hilbert space. We consider a given decomposition of a $d_A \otimes
d_B $ state $\varrho\in B(\mathcal{H}_A\otimes \mathcal{H}_B)$ and
construct an associated entanglement witness $W_{\varrho}$ acting
on a larger product Hilbert space ${\cal H}' \otimes {\cal H}'$
with $\dim \mathcal{H}'\leq (d_A d_B)^{4}$.
Weak optimality of this witness is then proven to be equivalent to
the separability of $\varrho$, where we call a witness weakly optimal
if its expectation value vanishes on at least one product vector,
or, in other words, it is tangent to the set of separable states
(see also Refs. \cite {Optimisation,ReviewOptimisationEtc} for the
notion of optimality of the entanglement witnesses).

Our approach has the following conceptual advantage: Since the
witness $W_{\varrho}$ can be explicitly calculated, all the
elements of the possible subsequent tests have well-defined and
clear structures. In particular, the {\it arbitrary multipartite}
separability problem is here mapped into the analysis of {\it a single
bipartite} entanglement witness (see Fig. 1). Moreover, our
formulation provokes some interesting questions about the
structure of the set of the entanglement witnesses and the
corresponding maps derived from a given quantum state.

\begin{figure}
\label{Fig2} \centering
\includegraphics[width=1\columnwidth]{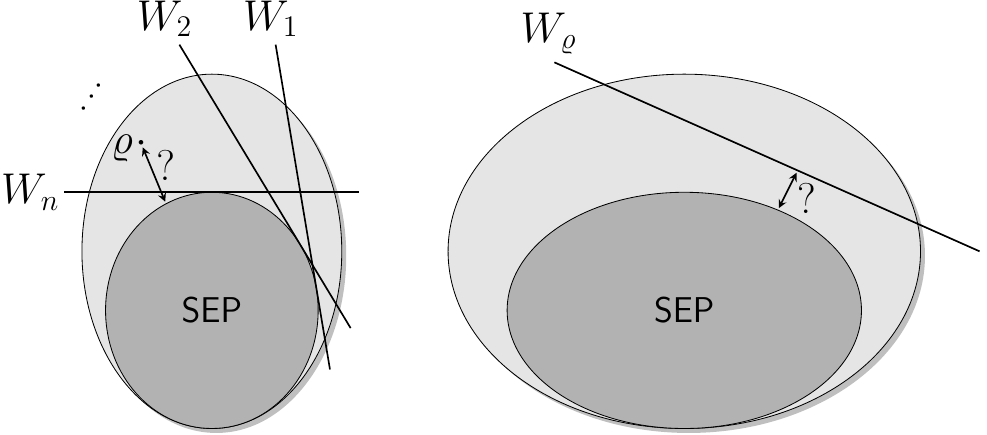}
\caption{The separability problem originally expressed in terms of
infinitely many entanglement witnesses (left) is here proved to be
equivalent to weak optimality of a single entanglement witness in
a larger Hilbert space (right).}
\end{figure}

In this context, it is worth noticing that the question of strict
positivity of a single entanglement witness on separable states
has an algorithmic solution in terms of the so-called Henkel
forms. The underlying algorithm was constructed more than three
decades ago by Jamio\l{}kowski \cite{Jam72forms} (see also Ref.
\cite{algorytmy}). Even though it is not of practical use here, it
is still conceptually interesting. In particular, 
this algorithm can decide the witness'
weak optimality in a finite, \textit{a priori known} number of
steps.

\textit{Construction and the main result.} Our state witness
$W_{\varrho}$ is constructed from the biconcurrence matrix
\cite{Badziag02}, two forms reflecting its transformation
properties and an additional projection. We begin the construction
with a decomposition of $\varrho$ in terms of subnormalized
vectors, so that $\varrho=\sum_{i} |\Psi^{i}\rangle\!\langle
\Psi^{i}|$ (eigendecomposition is usually the most obvious
although by no means necessary choice). The decomposition defines
the corresponding biconcurrence matrix $B=B(\varrho)$
\cite{Badziag02}, whose elements can most easily be expressed as
\cite{Mintert04}:
\begin{equation}
\hspace{-0.2cm}B_{m\mu,n\nu}=\langle \Psi^{m}_{AB}| \langle \Psi^{\mu}_{A'B'}|
P_{AA'}^{\mathrm{asym}} \otimes P_{BB'}^{\mathrm{asym}} | \Psi^{n}_{AB} \rangle|
\Psi^{\nu}_{A'B'} \rangle
\end{equation}
with $P^{\mathrm{asym}}_{XX'}$ being a projector onto the
antisymmetric subspace of the Hilbert space
$\mathcal{H}_X\ot\mathcal{H}_{X'}$ ($X=A,B$).

The operator $B$ acts on a product Hilbert space
$\mathcal{H}\ot\mathcal{H}$ with the dimension of $\mathcal{H}$ depending on the
amount of vectors $\ket{\Psi^i}$ in the above decomposition of $\varrho$. When
one begins with the eigendecomposition of $\varrho$, then
$\dim \mathcal{H}=d_{A}d_{B}$, which is the maximal number of eigenvectors of
$\varrho$. However, in order to allow for a separable decomposition (all
vectors $\ket{\Psi^i}$ are product) whenever it exists, one needs
$\dim\mathcal{H}=(d_Ad_B)^2\equiv N$ (recall that a separable
decomposition of $\varrho$ may require up to $N$ vectors).
Consequently, from now on we will regard $B$ as an operator acting
on the extended space ${\cal H} \otimes {\cal H}$ with $\dim {\cal
H}=N$. One then notices that $B$ is positive and symmetric with
respect to the transposition of indices $m$ and $\mu$, as well as
$n$ and $\nu$. More importantly, denoting by
$P_{\mathrm{cl}}=\sum_{i}\proj{i,i}$ the classically correlated
projector acting on $\mathcal{H}\ot\mathcal{H}$, it follows from
Ref. \cite{Badziag02} that $B$ is related to
separability of $\varrho$ via the following theorem.
\begin{thm}\label{thm1}
The state $\varrho\in B(\mathcal{H}_A\otimes\mathcal{H}_B)$ is
separable if and only if the function
\begin{equation}
\mathcal{B}(\varrho)=\inf_{U} \mathrm{Tr}\left(P_{\mathrm{cl}}U
\otimes U B U^{\dagger} \otimes U^{\dagger}\right), \label{Bfunc}
\end{equation}
called biconcurrence function, vanishes. The infimum is taken over
all unitary matrices $U$ acting on $\mathcal{H}$.
\end{thm}

As each unitary matrix in the above represents an orthonormal
basis, we can straightforwardly rewrite (\ref{Bfunc}) as

\begin{equation}
{\cal B}(\varrho)= \inf_{\{ |x_{i}\rangle \}} \sum_{i=1}^{N}
\langle x_{i} , x_{i}| B| x_{i},x_{i} \rangle=0
\label{g00},
\end{equation}
where the infimum is taken over all orthonormal bases $\{
|x_{i}\rangle \}$ of $\mathcal{H}$. Consequently, Theorem
\ref{thm1} can be alternatively phrased as follows.

\begin{thm}\label{thm2}
A bipartite state $\varrho$ is separable if and only if there
exists a set of vectors $| x_{i} \rangle, \ i=1 \ldots N$, for which
the following three forms vanish at the same time:
\begin{enumerate}[(i)]
\item zero form condition,
\begin{equation}
G_{0}=\sum_{i=1}^{N} \langle x_{i} , x_{i}| B| x_{i},x_{i} \rangle=0; \label{g0}
\end{equation}

\item orthogonality condition for the vectors $|x_{i}\rangle$,
\begin{eqnarray}
G_{1} &=&\sum_{\substack{i,j=1\\i\neq j}}^{N}|\langle x_i|x_j\rangle|^{2}=0;
\label{g1}
\end{eqnarray}
\item normalization condition for the basis $\{|x_{i}\rangle \}$,
\begin{eqnarray}
G_{2} &=&\sum_{i=1}^{N}\|x_i\|^4-\frac{1}{N}\left(\sum_{i=1}^{N}
\|x_i\|^2\right)^2=0.
\label{g2}
\end{eqnarray}
\end{enumerate}
\end{thm}

Since in general all $G_{0},G_{1},G_{2}$ are non-negative, these
three conditions can be replaced by a single one: $\alpha G_{0} +
\beta G_{1} + \gamma G_{2}=0$ for any fixed $\alpha, \beta, \gamma
> 0$. In other words, a state $\varrho$ is entangled if and only
if the inequality
\begin{equation}
\alpha G_{0} + \beta G_{1} + \gamma G_{2}>0 \label{g012}
\end{equation}
holds for any set of vectors $\{ |x_{i}\rangle \}$ and triple
$\alpha,\beta,\gamma>0$.

To convert this into the property (a) of an entanglement witness
(see above), we need to extend the Hilbert space once more. Recall
that the operator $B$ is defined on $\mathcal{H} \otimes
\mathcal{H}$. We extend each $\mathcal{H}$ to
$\mathcal{H}'=\mathcal{H}\otimes \widetilde{\mathcal{H}}$, where
$\widetilde{\mathcal{H}}$ is an auxiliary space isomorphic to
${\cal H}$. One then notices that any vector $|u\rangle \in
\mathcal{H}\otimes \widetilde{\mathcal{H}}$ can be written in the
form
\begin{equation}
|u\rangle=\sum_{i=1}^{N}|x_{i},i\rangle,
\end{equation}
where $\{\ket{x_i}\}$ is a set of arbitrary vectors from $\mathcal{H}$, while
$\{\ket{i}\}$ the standard basis in $\widetilde{\mathcal{H}}$.
This observation allows us to substitute single vectors in the
extended space for the sets of vectors in the conditions
(\ref{g0})--(\ref{g2}). To this end, let us introduce the swap
operator $V=\sum_{ij} |i\rangle\!\langle j|\ot |j \rangle\!
\langle i|$ that together with $P_{\mathrm{cl}}$ (see above for
the definition) will act on $\widetilde{\mathcal{H}} \otimes
\widetilde{\mathcal{H}}$. For the sake of clarity, we mark this
action by a tilde on top of the relevant operator.

With this notation, we can rewrite the necessary and sufficient
condition for entanglement (\ref{g012}) in terms of a degree-four form $A$
as
\begin{equation}
\mathcal{B}(\varrho)=\min_{\ket{u}\in\mathcal{H}\ot\widetilde{\mathcal{H}}}
\langle u,u| A| u,u
\rangle>0. \label{A}
\end{equation}
The minimum is taken over all vectors $|u \rangle \in
\mathcal{H} \otimes \widetilde{\mathcal{H}}$, while the operator
$A$ acts on $(\mathcal{H} \otimes \widetilde{\mathcal{H}}) \otimes
(\mathcal{H} \otimes \widetilde{\mathcal{H}})$ and reads:
\begin{eqnarray}\label{Amatrix}
A &=& \alpha B \otimes \widetilde{P}_{\mathrm{cl}} + \beta
I\otimes\big(\widetilde{V} -
\widetilde{P}_{\mathrm{cl}}\big)+ \gamma
I\otimes\Big(\widetilde{P}_{\mathrm{cl}} -
\widetilde{I}/N\Big)\nonumber\\
 &=& \alpha A_0 + \beta A_1 + \gamma A_2.
\end{eqnarray}
The parameters $\alpha,\beta,\gamma>0$ here can be chosen at will,
and this freedom may be utilized for, e.g., optimization of the
numerical separability tests based on condition (\ref{A}).

Each of the three terms contributing to the operator $A$ has
non-negative expectation values on symmetric product vectors
$\ket{u,u}$. Neither the whole operator nor any of its parts is,
however, a witness. On the one hand, the operator $A_0$ is
positive and clearly weakens the witness obtained using our method
(when the constructed witness is so weak that it is not even
weakly optimal, then the corresponding state $\varrho$ is
entangled). On the other hand, the operators $A_1$ and $A_2$ do
not represent entanglement witnesses since they have negative
expectation values on some product vectors $|u, v \rangle $ with
$\ket{u}\neq \ket{v}$. One can, nevertheless, remove this
disadvantage without affecting the expectation values on symmetric
product vectors $|u, u \rangle $ by adding to $A$ a projection on
the antisymmetric subspace $P^{\mathrm{asym}}= (1/2)(I \ot
\widetilde{I} - V \ot \widetilde{V})$ with large enough weight.
Moreover, without affecting the expectation values $\langle
u,u|A|u,u \rangle$ in (\ref{A}), it is possible to substitute $Y =
P^{\mathrm{sym}} A P^{\mathrm{sym}}$ for the original operator
$A$. When one has done the latter, then the following lemma gives
a straightforward method to calculate a weight with which
$P^{\mathrm{asym}}$ has to be added to an operator $A_1$ or $A_2$
to guarantee its conversion into an entanglement witness.
\begin{lem}\label{lemat}
Let $X$ be a Hermitian operator acting on a product Hilbert space
$\mathcal{H} \otimes \mathcal{H}$ such that
$X=P^{\mathrm{sym}} X
P^{\mathrm{sym}}$ and $\langle u, u|X |u ,u \rangle \geq 0$ for
any $\ket{u}\in\mathcal{H}$. Moreover, let $X_{C}=X +
CP^{\mathrm{asym}}$, where  $P^{\mathrm{asym}}$ projects onto the
antisymmetric subspace of $\mathcal{H} \otimes \mathcal{H}$ and $C$ is
a real constant. Then the following implications are true:
\begin{enumerate}[(i)]
\item If $C\geq
\|X\|_{\infty}$, then $\langle u,v|X_{C}|u,v\rangle\geq 0$
holds for any pair of vectors $\ket{u},\ket{v}\in\mathcal{H}$,

\item if $C\geq 2 \|X\|_{\infty}$, then for any
pair of vectors $\ket{u},\ket{v}\in\mathcal{H}$,
there exists $\ket{g}\in\mathcal{H}$ such that
\begin{eqnarray}\label{12}
\langle u, v| X_{C} |u, v \rangle &\geq& \langle
g, g| X_{C} |g, g \rangle\nonumber\\
&\geq& \inf_{\ket{u}\in\mathcal{H}} \langle u, u| X |u, u \rangle
\;(=:\mathcal{X}).
\end{eqnarray}
\end{enumerate}
\end{lem}
\begin{proof}
Taking two arbitrary normalized vectors $\ket{u},
\ket{v}\in\mathcal{H}$, the symmetry $X=P^{\mathrm{sym}} X
P^{\mathrm{sym}}$ implies that
\be
\langle u,v| X |u,v \rangle= \langle \Psi| X
|\Psi\rangle, \label{XPsi}
\ee
where $|\Psi \rangle :=(|u,v\rangle+|v,u\rangle)/2$. Up to an
unimportant global phase factor, the vector $|v\rangle$ can be
decomposed as $|v\rangle=a| u\rangle + b |u^{\perp}\rangle$ with
$a,b$ being two non-negative numbers such that $a^{2} +b^{2}=1$ and
$\ket{u^{\perp}}$ denoting a vector orthogonal to $\ket{u}$.
Consequently, $|\Psi \rangle =a |u\rangle|u\rangle + b (|u\rangle
|u^{\perp}\rangle + |u^{\perp}\rangle |u\rangle)/2 $ and
$\|\Psi\|^{2}=a^{2} + b^{2}/2$. Finally, it is fairly easy to
convince oneself that the Schmidt decomposition of $|\Psi \rangle$
reads $|\Psi \rangle = x |e,e\rangle + y |f,f\rangle$, where
$\ket{e}=[1/\sqrt{2(1+a)}][(1+a)\ket{u}+b\ket{u^{\perp}}]$ and
$\ket{f}=[\mathrm{i}/\sqrt{2(1-a)}][(1-a)\ket{u}-b\ket{u^{\perp}}]$
are orthonormal vectors, while $x=(1+a)/2$ and $y=(1-a)/2$. All
this allows us to write that
\begin{eqnarray}\label{estimation}
\langle u, v| X |u , v \rangle&\!\!=\!\!& \langle \Psi| X |\Psi\rangle \nonumber
\\
&=&x^{2} \langle e,e| X |e, e \rangle + y^{2} \langle f, f| X |f,f \rangle
\nonumber \\
&& + 2xy \mathrm{Re}(\langle e, e| X |f, f \rangle) \nonumber \\
&\geq& x^{2} \langle e, e| X |e,e \rangle + y^{2} \langle f, f| X |f,f \rangle
\nonumber \\
&&- 2xy |\langle e, e| X |f, f \rangle| \nonumber \\
&\geq& x^{2} \langle e, e| X |e, e \rangle + y^{2} \langle f, f| X |f, f
\rangle-2xy C \nonumber \\
&\geq&-\tfrac{1}{2}(1-a^{2})C=-\tfrac{1}{2}(1-|\langle u|v\rangle|^{2})C
\nonumber \\
&=&-C \langle u,v|P^{\mathrm{asym}}|u,v \rangle,
\label{inequalities}
\end{eqnarray}
where the first inequality follows from the fact that
$\mathrm{Re}z\leq |z|$ holds for any $z\in\mathbbm{C}$, while the
second and the third ones from the assumptions that, respectively,
$C\geq\|X\|_{\infty}$ and $\langle u, u|X |u,u \rangle \geq 0$ for
any $\ket{u}$.

Comparison of the first and the last expression in
(\ref{inequalities}) immediately gives
%
$\langle u, v| X + CP^{\mathrm{asym}}|u, v \rangle\geq 0 $
%
for all $\ket{u},\ket{v}\in\mathcal{H}$ and $C \geq
\|X\|_{\infty}$, proving (i).

In order to prove (ii), we can exploit the second inequality in
(\ref{inequalities}). Its right-hand side does not exceed $(x^{2}
+y^{2}) \langle \tilde{e}, \tilde{e}| X |\tilde{e},\tilde{e}
\rangle - 2xyC$, where
$\ket{\tilde{e}}=\ket{e}$ if
$\bra{e,e}X\ket{e,e}\leq\bra{f,f}X\ket{f,f}$, and $\ket{\tilde{e}}=\ket{f}$
otherwise. Consequently,
\be
\langle u, v| X |u , v \rangle
\geq (x^{2} +y^{2}) \langle \tilde{e},\tilde{e}| X
|\tilde{e}, \tilde{e} \rangle - 2xy C
\ee
which can be rewritten as
\be
\langle u, v| X |u , v \rangle  +
4xyC\geq (x^{2} +y^{2}) \langle \tilde{e},\tilde{e}| X
|\tilde{e},\tilde{e} \rangle +2xy C. \label{laststep}
\ee
Utilizing further the fact that $C \geq \langle \tilde{e},
\tilde{e}| X |\tilde{e}, \tilde{e} \rangle $ on the right-hand
side of (\ref{laststep}), we arrive at
\be
\langle u, v| X |u,v \rangle  + 4xyC\geq (x+y)^{2} \langle \tilde{e},
\tilde{e}| X |\tilde{e}, \tilde{e} \rangle,
\ee
which, due to the facts that $x+y=1$ and $4xy=2\langle u, v
|P^{\mathrm{asym}}|u, v \rangle$, simplifies to
\be
\langle
u,v| X +2C P^{\mathrm{asym}}|u ,v \rangle \geq  \langle
\tilde{e},\tilde{e}| X |\tilde{e},\tilde{e}
\rangle.
\ee
After replacing $2C$ by $C$ and using the assumption that $C
\geq2\|X\|_\infty$, this finally gives (\ref{12}), concluding the
proof.
\end{proof}

Let us notice that the property (ii) implies in particular that
$\mathcal{X}_{C}:=\inf_{\ket{u},\ket{v}} \langle u, v| X_{C} |u, v
\rangle \nonumber = \mathcal{X}$. In other words, if we choose a
sufficiently large $C$, then the expectation value of $X_{C}$ in a
separable state always upper bounds $\mathcal{X}$.

Our matrix $Y$ satisfies the assumptions of the lemma.
Consequently $Y_{C}=Y  + CP^{\mathrm{asym}}$ ($C \geq
2\|Y\|_{\infty}$) is a good candidate for an entanglement witness.
In fact, it is a witness, since it has at least one negative
eigenvalue. In this way we have arrived at the main result of the
paper.

\begin{thm}\label{thm_central}
A bipartite state $\varrho$ is separable if and only if its
corresponding entanglement witness $W_{\varrho}=Y_{C}$  with $C >
\|Y\|_{\infty}$ is weakly optimal. Moreover, if $C \geq
2\|Y\|_{\infty}$, then the witness satisfies in addition the
condition (\ref{12}), guaranteeing that
$\bra{u,v}W_{\varrho}\ket{u,v}\geq \mathcal{B}(\varrho)$ for all
$\ket{u}$, $\ket{v}$.
\end{thm}
\begin{proof}First, (i) of Lemma \ref{lemat} guarantees that for any
$C\geq \|Y\|_{\infty}$, $W_{\varrho}$ is an entanglement witness. Then, it
follows from the estimation (\ref{estimation}) that if $C>\|Y\|_{\infty}$,
$\langle u,v|W_{\varrho}|u,v\rangle >0$ for all $\ket{u}\neq \ket{v}$, meaning
that the witness $W_{\varrho}=Y_C$ can be tangent to the set of separable states
only on the symmetric product vectors $\ket{u,u}$. This, in view of Theorem
\ref{thm2} and the discussion that follows, means that the state $\varrho$ is
separable if and only if the corresponding witness $W_{\varrho}$ is weakly
optimal. It should be noticed that if $W_{\varrho}$ is weakly optimal for some
$C>\|Y\|_{\infty}$ then it is weakly optimal for any such $C$.
To prove the second part of the theorem one combines (ii) of
Lemma \ref{lemat} and (\ref{A}).
\end{proof}

A simple corollary to this theorem provides a direct link between
separable states from $B(\mathcal{H}_A\ot\mathcal{H}_B)$ and
weakly optimal entanglement witnesses acting on
$\mathcal{H}'\ot\mathcal{H}'\cong\mathbbm{C}^{N^2} \otimes
\mathbbm{C}^{N^2}$ with $N=(d_Ad_B)^2$, namely:

\begin{cor}
Every separable state with a pure state product decomposition of
length $N$  generates a corresponding weakly optimal entanglement
witness from $B(\mathbbm{C}^{N^2} \otimes \mathbbm{C}^{N^2})$.
\end{cor}

Clearly, the strongest entanglement witnesses constructed in this
way are those for $A_0=0$. Even then, however, the witness
construction based on Lemma \ref{lemat}, although universal, does
not have to produce the most interesting witnesses.
To illustrate this point, we consider the choice $\beta=\gamma=1$
and put $A_0 = 0$. The resulting operator (\ref{Amatrix}) is then
$A_{12} = I \ot \widetilde{V} - (1/N) I \ot \widetilde{I}$, while
its symmetrization reads $Y = (1/2)[I \ot \widetilde{V} + V \ot
\widetilde{I} - (1/N) (I \ot \widetilde{I} + V \ot
\widetilde{V})]$. With a little bit of work, one can easily check
that $\left\| Y \right\|_{\infty} = (N+1)/N$. According to Lemma
\ref{lemat}, one then needs to add $[(N+1)/N] P^{\mathrm{asym}}$
to $Y$, in order to secure its conversion into an entanglement
witness $W^{\mathrm{sym}}$. Apparently, this is quite unnecessary.
Knowing that for any operator $X \in B(\mathbbm{C} ^n)$, $\left\|
X \right\|_{\mathrm{Tr}} \leq \sqrt{n} \left\| X
\right\|_{\mathrm{HS}}$ ($\left\| \cdot \right\|_{\mathrm{Tr}}$
and $\left\| \cdot \right\|_{\mathrm{HS}}$ stand for,
respectively, the trace and the Hilbert-Schmidt norm), one can
easily show that without any symmetrization, it is enough to add
$(2/N) P^{\mathrm{asym}}$ to $A_{12}$ in order to convert it into
a witness operator $W = I \ot \widetilde{V} - (1/N) V \ot
\widetilde{V}$. It follows that $W$ belongs to the class of the
so-called decomposable witnesses (see Ref. \cite{Optimisation}).
Witnesses as $W^{\mathrm{sym}}$ and $W$ may still have zero
expectation values on some product vectors $\left|uv\right\rangle$
with $\ket{u} \neq \ket{v}$. For that, they do not make any good
ground for entanglement identification in $\varrho$. To remedy
this disadvantage, it is, however, enough to add
$P^{\mathrm{asym}}$ with any positive weight to these witnesses
(see the comment after theorem \ref{thm_central}). While this will
not change their expectation values on symmetric product vectors
$\left|uu\right\rangle$, the new witnesses (let us denote them by
$W_{+}^{\mathrm{sym}}$ and $W_{+}$) will become strictly positive
on all products $\left|uv\right\rangle$ with $\ket{u} \neq
\ket{v}$. This is enough to guarantee that after the addition of
the contribution from $A_0$, the resulting witness will be weakly
optimal if and only if the state $\varrho$, from which $A_0$ (via $B$) is
derived, is separable \cite{Przypis}.

Our method of linking separability of a bipartite state to weak
optimality of a single entanglement witness readily generalizes
for the states shared by many parties. In the latter case,
however, different aspects of separability are described by
different matrices $B$ \cite{Mintert05}. Thus, one will end up
with different corresponding operators $A_0$, depending on which
aspect of multi-partite entanglement (separability) one would like
to test. Nevertheless, the design and structure of the
state-independent contributions to our witness ($A_1$ and $A_2$)
as well as condition (\ref{A}), together with (\ref{Amatrix}),
will be exactly as in the bipartite case, irrespectively of the
number of parties sharing the tested state $\varrho$.
Consequently, the design and the properties of $W_{\varrho}$ for a
multi-partite $\varrho$ will be exactly the same as in the
bipartite case.

\textit{Connection to the theory of positive maps.} Via the
Jamio\l{}kowski isomorphism, the relation between bipartite states
and their ``state witnesses'' directly translates into a relation
between bipartite states and positive but not completely positive
maps. In particular, it is easy to see that in the isomorphism,
operators, which are not weakly optimal, are mapped onto fully
mixing maps. These are the maps which transform any state into a
positive matrix of full rank. We then have another immediate
corollary to Theorem~\ref{thm_central}.

\begin{cor}
A bipartite state $\varrho$ is entangled  if and only if a
positive map $\Lambda_{\varrho}$ (it can be chosen to be
decomposable) corresponding through the Jamio\l{}kowski
isomorphism to the witness $W_{\varrho}$ is fully mixing.
\end{cor}

Indeed, the choice of parameters ($\beta = \gamma$) produces
decomposable witnesses and thus decomposable maps.

\textit{Conclusion.} The separability problem is known to be
computationally hard \cite{Gurvits}. Nevertheless, analysis of the
properties of  witnesses $W_{\varrho}$ (and maps
$\Lambda_{\varrho}$) should be at least in some cases relatively
straightforward. One can then hope that our approach not only
sheds light on the conceptual aspect of the separability
problem, but also may become a starting point for the development of
more efficient numerical separability tests. Finally,
allowing for $\beta \neq \gamma$ in formula (\ref{Amatrix}) may
lead to nondecomposable witnesses and nondecomposable maps. This
in turn may lead to some questions about the nature of these
witnesses, their possible relation to potential bound entanglement
in $\varrho$, or their ability to reveal different geometrical properties
of the boundary of the set of separable states. We leave these
questions for further research.

\textit{Acknowledgements.} We thank J. Eisert, M. Demianowicz, L.
Ioannou, C. Mora and M. Piani for fruitful discussions. This work
is supported by  Polish Ministry of Science and Education under
Grant No. 1 P03B 095 29 and EU project SCALA FP6-2004-IST
No. 015714.


\begin{thebibliography}{0}

\bibitem{Peres}A. Peres, Phys. Rev. Lett. {\bf 77}, 1413 (1996).

\bibitem{sep}M. Horodecki, P. Horodecki, and R. Horodecki  Phys. Lett. A
{\bf 223}, 1 (1996).

\bibitem{JamiolkowskiIsomorphism}
A. Jamio\l{}kowski, Rep. Math. Phys. {\bf 3}, 275 (1972).

\bibitem{Terhal}B. Terhal, Phys. Lett. A {\bf 271}, 319 (2000).

\bibitem{Exp}B. Terhal, Linear Algebra Appl. {\bf 323}, 61 (2000).



\bibitem{Witn1}F. G. S. L. Brand\~{a}o, Phys. Rev. A \textbf{72}, 022310 (2005);
K. M. R. Audenaert and M. B. Plenio, New J. Phys. {\bf 8}, 266 (2006);
J. Eisert, F. G. S. L. Brand\~{a}o, and K. M. R. Audenaert, \textit{ibid.} {\bf
9}, 46 (2007); O. G\"uhne, M. Reimpell, and R. F. Werner, Phys. Rev. Lett.
\textbf{98}, 110502 (2007); Phys. Rev. A \textbf{77}, 052317 (2008).


\bibitem{CovarianceSep}
O. G\"uhne {\it et al.}, Phys. Rev. Lett. \textbf{99}, 130504
(2007); J. Samsonowicz, M. Ku\'s, and M. Lewenstein, Phys. Rev. A
\textbf{76}, 022314 (2007);
M. Seevinck and J. Uffink, \textit{ibid.} \textbf{78}, 032101 (2008);
T. Moroder, O. G\"uhne, and N. L\"utkenhaus, \textit{ibid.} \textbf{78}, 032326
(2008);
R. Augusiak and J. Stasi\'nska, New J. Phys. \textbf{11}, 053018
(2009); O. G\"uhne and M. Seevinck, \textit{ibid.} 12, 053002
(2010).


\bibitem{przegladowa}R. Horodecki {\it et al.},
Rev. Mod. Phys. \textbf{81}, 865 (2009).

\bibitem{przeglDet}O. G\"uhne and G. T\'oth, Phys. Rep. \textbf{474}, 1 (2009).
    
\bibitem{Optimisation}
M. Lewenstein {\it et al.}, Phys. Rev. A {\bf 62}, 052310 (2000).

\bibitem{swiadki:wlasn}M. Lewenstein, B. Kraus, P. Horodecki,
and J. I. Cirac, Phys. Rev. A \textbf{63}, 044304 (2001); J. K.
Korbicz \textit{et al.}, \textit{ibid.} \textbf{78}, 062105 (2008);
G. Sarbicki, J. Phys. A \textbf{41}, 375303 (2008); R. Augusiak,
J. Tura, and M. Lewenstein, \textit{ibid.} \textbf{44}, 212001 (2011);
D. Chru\'sci\'nski and G. Sarbicki, \textit{ibid.}
\textbf{45}, 115304 (2012); K.-C. Ha and S.-H. Kye, J. Math. Phys.
\textbf{53}, 102204 (2012).


\bibitem{Experiments}
M. Barbieri {\it et. al.}, Phys. Rev. Lett. {\bf 91}, 227901
(2003); M. Bourennane {\it et al.}, {\it ibid.} {\bf 92}, 087902
(2004); K. J. Resch, P. Walther, and A. Zeilinger, {\it ibid.}
{\bf 94}, 070402 (2005); J. Altepeter {\it et al.}, {\it ibid.}
{\bf 95}, 033601 (2005); N. Kiesel {\it et. al.}, {\it ibid.} {\bf
95}, 210502 (2005); H. H\"affner {\it et al.}, Nature (London) {\bf 438},
643 (2005); C.-Y. Lu \textit{et al.}, Nat. Phys. \textbf{3}, 91
(2007).


\bibitem{ReviewOptimisationEtc}
D. Bruss {\it et al.}, J. Mod. Opt. {\bf 49}, 1399 (2002).

\bibitem{Jam72forms}
A. Jamio\l{}kowski, \textit{An Effective Method for Investigation of
Positive Endomorphisms on the Set of Positive Definite Operators},
Nicolaus Copernicus University (Toru\'n, Poland) Report No. 175, 1972
(unpublished).

\bibitem{algorytmy}G. Dahl, J. M. Leinaas, J. Myrheim, and E. Ovrum,
Linear Algebra Appl. \textbf{420}, 711 (2007);
J. Sperling and W. Vogel, Phys. Rev. A \textbf{79}, 022318 (2009);
J. M. Leinaas, J. Myrheim, and P. \O. Sollid, \textit{ibid.} \textbf{81}, 062329
(2010).

\bibitem{Badziag02}
P. Badzi{\c{a}}g {\it et al.}, J. Mod. Opt. {\bf 49}, 1289 (2002).



\bibitem{Mintert04}
F. Mintert, M. Ku\'s, A. Buchleitner, Phys. Rev. Lett.
{\bf 92}, 167902 (2004).

\bibitem{Przypis}
Note that in this case one can apply Lemma \ref{lemat} to the two
parts (witnesses) $\alpha A_{0}$ and $\beta A_{1} + \gamma A_{2}$
(equal to $W$ or $W^{\mathrm{sym}}$, respectively) independently.
The only effort required here to get the condition (\ref{12}) is
to calculate the norm of matrix $B$ since the norms of $W$,
$W^{\mathrm{sym}}$ are trivially computable.

\bibitem{Mintert05}
F. Mintert, M. Ku\'s, A. Buchleitner, Phys. Rev. Lett.
{\bf 95}, 260502 (2005).



\bibitem{Gurvits}L. Gurvits, in \textit{Proceedings of the Thirty-Fifth Annual
ACM Symposium on Theory of Computing (STOC’03), San Diego,
2003} (ACM, New York, USA, 2003), pp. 10–19.




\end{thebibliography}
\end{document}